\documentclass[12pt,aps,prb,preprint]{revtex4}   % style for Physical Review B and AJP are similar

\usepackage{amsmath}    % need for subequations
\usepackage{graphicx}   % for figures

\draft
\begin{document}
\newcommand{\la}{{\langle}}
\newcommand{\ra}{{\rangle}}
\newcommand{\eps}{{\epsilon}}
\newcommand{\lam}{{\lambda}}
\newcommand{\del}{{\delta}}
\newcommand{\Del}{{\Delta}}
\newcommand{\eq}{{\equiv}}
\newcommand{\nn}{{\nonumber}}
\newcommand{\Ga}{{\Gamma}}
\title{A General Criterion of Quantum Integrability Accommodating Central Charges and "Anomalies"}
%Lines break automatically or can be forced with \\

\author{P.P.Divakaran}
% \altaffiliation[Also at ]{home.}  %  optional
 \affiliation{The Institute of Mathematical Sciences, CIT Campus, Chennai 600 113, India}
% \email{hgould@clarku.edu}   %optional

\begin{abstract}
A simple quantum generalisation of the Liouville--Arnold criterion 
of classical integrability is proposed: a system is quantum-integrable
if it has an abelian Lie group of Wigner symmetries of dimension 
equal to the number of degrees of freedom. The criterion goes 
significantly beyond the familiar case of involutive conserved 
operators to cover 
systems with anomalies, in which involutivity is modified by central 
charges. "Anomalous" quantum integrability is shown to have all the expected consequences including exact 
diagonalisability.The approach throws new light on the origin of Weyl group 
invariance.
\end{abstract}

\maketitle
1.{\it Classical and quantum integrability}. A constraint-free conservative mechanical system of a finite number $n$ of degrees of freedom is said to be completely integrable classically (abbreviated to LA-integrable, for Liouville-Arnold)[1]  if there are $n$ functions ${f_i}$ on its phase space $P$ having the properties of (i)
involutivity: ${f_i}$ have vanishing Poisson brackets: $[f_i,f_j]=0$; (ii) conservation: $[h,f_i]=0$ where $h$ is the Hamiltonian function; and (iii)completeness: the differentials ${df_i}$ are linearly independent. The conserved quantities ${f_i}$ are action variables and they form one half of a set of canonical coordinates ${f_i,q_i}$ for $P$, $[f_i,q_j]=\delta_{ij}$. More precisely, ${q_i}$ are local coordinates for a maximal submanifold $C$ (of dimension $n$) of $P$ on which all $f_i$ take constant values. $C$ is a generalised configuration space adapted to the set of action variables. It is not necessarily compact; despite this, we shall refer to $q_i$ as an angle variable. It is an immediate consequence of LA-integrability that $h$ is a function only of ${f_i}$ and that the equation of motion for $q_i$ takes the form $dq_i/dt=\phi_i({f})$ for some function $\phi_i$ and so can be trivially integrated.

Our purpose here is to propose an answer to the question: What is a criterion for quantum (q-)integrability that is as concise and general as the above classical criterion? Historically, the study of q-integrability began with certain soluble models[2,3] exhibiting the properties (i) to (iii), with obvious reinterpretations of ${f}$ and $h$ as selfadjoint operators on the space of states and of [ , ] as the commutator. (We shall later be using [ , ] for the Lie brackets also; the context will make the meaning clear). That these properties are a satisfactory general characterisation of q-integrability seems to have general acceptance, though their verification in models remains a case by case exercise. In addition to having diagonalisable Hamiltonians, such models exhibit certain finite(Weyl) group "symmetries"[3,4] whose provenance has remained a matter of some mystery.

What is proposed here 
is an inherently quantum formulation of q-integrability that relies on the classical theory only for 
motivation. It provides a framework covering the models mentioned above, namely those obtained by a direct transcription 
("naive quantisation") of an integrable classical system to the quantum domain. But it also extends the 
scope of the notion, beyond such "normal" q-integrable systems, to a class of "anomalous" q-integrable 
systems for which involutivity is modified by the presence of central charges. (These terms are explained 
later).In particular, the latter class will be shown to possess the properties of diagonalisability 
and Weyl group invariance.

2.{\it The criterion}.The standard criterion of LA-integrability stated 
above says that the system has an $n$-dimensional abelian Lie algebra of 
classical symmetries. We shall assume that this can be exponentiated 
to a Lie group $G$ (connected, abelian, $n$-dimensional) of symmetries. 
Classically, this only requires that $P$ is a manifold[5]; in the 
quantum context, symmetry under $G$ is a tighter demand than under its Lie 
algebra. Our criterion for q-integrability is then simply:

{\it A quantum system of $n$ degrees of freedom is q-integrable if it 
has a connected, abelian, $n$-dimensional Lie group $G$ as a group of 
Wigner symmetries}. 

By a Wigner symmetry is meant as usual a one-one onto map of states to themselves preserving transition probabilities. Significantly, $G$-invariance of the Hamiltonian is not part of this criterion. Indeed demanding it will 
prove to be unjustified in general---we will see below that it is the 
proper handling of this issue that extends the scope of q-integrability 
to systems "with anomalies".

The group $G$ is the product of a vector group and a torus group: $G={\bf R}^k\times{\bf T}^{n-k}$, for some $k$, $ 0\leq k\leq n$. By Wigner's theorem, 
the state space ${\cal H}$ must carry (continuous) projective unitary
 representations (PURs in short) of this group. We summarise the 
  facts concerning them, relevant to the present work, briefly[6,7].

PURs of the group ${\bf R}^n$ are in one-one correspondence with, and can 
be obtained {\it via} the exponential map from, the URs of a Lie 
algebra $g_\alpha$  with basis $X_i$ and brackets 
$$
[X_i,X_j]=i\alpha_{ij},\hskip 5mm  1\leq i,j\leq n, $$
obtained by adjoining a set of real central charges $\alpha_{ij}=-\alpha_{ji}$ to the abelian Lie algebra of ${\bf R}^n$. These brackets define a central 
extension of ${\bf R}^n$, as a Lie algebra, by ${\bf R}$ and 
corresponding to each distinct set $\alpha=\{\alpha_{ij}\}$ we get a distinct central extension by $U(1)$, and a distinct equivalence class of PURs,
of the group ${\bf R}^n$. The representation space ${\cal {H}_\alpha}$
is a superselection sector in the total state space ${\cal {H}}$ for a fixed $\alpha$ and the latter is therefore a collection of sectors with no 
inter-sector transitions allowed, rather than their direct sum. The 
trivial sector ${\cal H}_0$ is special; it carries a UR (or a trivial 
PUR) of ${\bf R}^n$.

In sharp contrast, a torus group has only trivial central extensions 
and trivial PURs[8] even though ${\bf T}^n$ and ${\bf R}^n$ have the 
same Lie algebra; the state space consists of just the trivial sector. 
Thus the possible central charges are determined by the compactness 
properties of $G$.

3.{\it Normal q-integrability; the trivial sector}. We consider first the case $G={\bf T}^n$. Then the total state space ${\cal H}$ itself is the trivial sector; it carries a UR $U$ 
of ${\bf T}^n$ and hence of its Lie algebra ($\cong {\bf R}^n$) with 
basis $X_i$. We can choose the $n$ independent selfadjoint
operators $F_i=U(X_i)$ as action operators on ${\cal {H}}$ and they are 
trivially involutive. Next, we note that the conservation of $F_i$ 
is equivalent to the ${\bf T}^n$-invariance of the Hamiltonian operator $H$ of the system, by virtue of the equation of motion
$$
i\dot F_i=[H,F_i]=0.$$
But the invariance of $H$ follows from the fact that ${\bf T}^n$ acts unitarily on all states in ${\cal H}$  
at all times; thus conservation is ensured by our criterion. Consequently, as in the classical case, $H$ is a function of $F_i$ 
alone. To show this, it is best to choose the concrete realisation of 
${\cal {H}}$ as $L^2(C)$, the Hilbert space of square-integrable functions $\psi$ (generalised wave functions)   
on the generalised configuration space $C$ of angle variables $q_i$. 
The action 
$$
(F_i\psi)(q)=-i\frac{\partial\psi}{\partial q_i}$$ 
makes $L^2(C)$ a UR space of ${\bf T}^n$ (imposing suitable boundary conditions if needed). We have also the "angle operators"
$$
(Q_i\psi)(q)=q_i\psi(q),$$ so that on $L^2(C)$ there is an irreducible (by the Stone--von Neumann theorem) UR of the 
Heisenberg algebra defined by the canonical commutators $[F_i,Q_j]=-i\delta_{ij}$. Every vector of ${\cal H}$ can then be approximated by polynomials 
in the creation operators $F_i+iQ_i$ operating on the unique state annihilated by 
 $F_i-iQ_i$ for all $i$. Hence every densely defined operator, in particular $H$, can be 
approximated by polynomials in ("is a function of") $F_i,Q_i$, 
implying $[H,F_i]=i\partial H/\partial Q_i=0$ from the conservation 
of $F_i$ (invariance of $H$).

Thus normal q-integrability, which is a special case of our general 
criterion, is a direct quantum transcription of LA-integrability in all respects. The 
models of [2,3] are examples of this class. Their LA-integrability 
guarantees their normal q-integrability and {\it vice versa}; stated 
differently, every LA-integrable system has a normal q-integrable 
quantisation whose classical limit it is. It is also possible, from 
our quantum point of view, to determine the circumstances under which 
such systems possess finite (Weyl) group symmetries; this will be done elsewhere.

4.{\it Nontrivial sectors}. The fundamental difference between the 
trivial and a nontrivial sector is best seen in the maximally nontrivial 
case, namely, when the central charges form a nonsingular matrix $A$, 
det$A$ = 0. This means that $G={\bf R}^n$ with no torus factor and 
that $n=2l$ is even. ${\cal {H}}_\alpha$ can still be taken as the space 
$L^2(C)$ of generalised wave functions (many-valued wave functions may 
have to be considered, but this is easily done) with the angle operators 
acting, as before, by multiplication by $q_i$. As for the action operators, it is quickly checked 
that 
$$
F_{\alpha i}=U_\alpha(X_i)=-i\frac{\partial}{\partial q_i}+\frac{1}{2}\alpha_{ij}q_j $$
is a UR of the Lie algebra $g_\alpha$:
$$ [F_{\alpha i}, F_{\alpha j}] = i\alpha_{ij},$$
while preserving the action--angle CCR. This UR is not irreducible. 
The operators
$$
F'_{\alpha i}=-i\frac{\partial}{\partial q_i}-\frac{1}{2}\alpha_{ij}q_j $$
satisfy
$$
[F'_{\alpha i}, F_{\alpha j}]=0,\hskip 5mm [F'_{\alpha i}, F'_{\alpha j}]=-i\alpha_{ij}, $$
showing that ${\cal {H}}_\alpha$ carries a UR of the direct sum $g_\alpha\oplus g_{-\alpha}$. 
Hence ${\cal {H}}_\alpha$ has the tensor factorisation
$$
{\cal {H}}_\alpha={\cal {V}}_\alpha\otimes{\cal {V}}_{-\alpha},$$
where ${\cal {V}}_\alpha$ is the unique (upto equivalence, by Stone--von Neumann) irreducible (by Schur) UR 
of the Heisenberg Lie algebra $g_\alpha$. 

This factorisation[9] is a key result. It suggests that the operators best adapted to the study of 
q-integrability in the presence of central charges are not the action and angle operators but 
rather the action operators for both signs of the central charges. Note also that $g_{-\alpha}$ 
is not part of the integrability Lie algebra; it just describes concisely the (infinite) 
multiplicity in ${\cal {H}}_\alpha$ of the unique irreducible UR of $g_\alpha$. 

5.{\it Conservation, anomalies}. The superselection structure requires the Hamiltonian, like all 
observables, to be block-diagonal with respect to the sectors---it is actually a family of operators 
$H_\alpha$, one for each sector. Now the group unitarily represented in ${\cal H}_\alpha$ is not $G$ 
but rather its central extension $G_\alpha$ (of which $G$ is {\it not} a subgroup except for 
$\alpha=0$). Thus for each $\alpha$, $H_\alpha$ must be invariant under the unitary action 
of $G_\alpha$ on ${\cal H}_\alpha$. Infinitesimally, this demand generalises the trivial sector equation of 
motion (Sec.3) to 
$$
i\dot F_{\alpha i}=[H_\alpha,F_{\alpha i}]=0, $$
valid in an arbitrary sector. Thus all $F_{\alpha i},$ $i=1,..,2l,$ are necessarily conserved simply as a 
consequence of the criterion of q-integrability; conservation in ${\cal H}_\alpha$ and the $g_\alpha$-
invariance of $H_\alpha$ are, once again, the same. Contrarily, $F'_{\alpha i}$ are not conserved 
as they do not represent the symmetries of integrability.

We can now use the factorisation property of ${\cal H}_\alpha$ to show that $H_\alpha$ is {\it not} a function of $F_{\alpha i}$, 
but only of $F'_{\alpha i}$: briefly, from the irreducibility of ${\cal H}_\alpha$ under $g_\alpha\oplus
g_{-\alpha}$, $H_\alpha$ is {\it a priori} a polynomial in $F_{\alpha i}$ and $F'_{\alpha i}$ but, since it cmmutes with $F_{\alpha i}$, it 
cannot depend on them. The argument, based on properties of the Heisenberg Lie algebra, is 
the same as used for the trivial sector Hamiltonian $H$, but the conclusion is diametrically 
different. (Of course, for $\alpha=0$, $F_i$ and $F'_i$ coincide). We also note the implication 
that $H_\alpha$ cannot be chosen as one of the conserved charges $F_{\alpha i}$, as in normal 
q- (and LA-) integrable systems.

Given the generality of scope of our discussion, invariance is about the only guiding principle
available in choosing a Hamiltonian---starting with a system of partcles in a real configuration 
space with a specified Hamiltonian and transforming to action--angle operators to exhibit integrability 
is not part of our present aim. We can only assert that when the Hamiltonian is so transformed, 
it should not depend on the action operators if the system is integrable. In the rest of this 
paper we make a specific choice for $H_\alpha$ as a polynomial in $F'_{\alpha i}$, namely,
$$
H_\alpha = \sum_{i=1}^{2l}(F'_{\alpha i})^2.$$ 
This is the Hamiltonian operator in ${\cal H}_\alpha$ resulting from the appropriate quantisation 
of the classical Hamiltonian function $h=\Sigma_i(f_i)^2.$ 
The details below are for this particularly nice Hamiltonian, but similar results hold and can be worked 
out for other acceptable choices. 

On the other hand, a "naive"
quantisation (see Sec. 1) of the same system will lead to the choice $H=H_0=\Sigma_i(F_i)^2$ for commuting
 $F_i$, which is correct only for the trivial sector. Using this $H_0$ to compute time evolution
in ${\cal H}_\alpha$ results in an apparent lack of conservation of $F_{\alpha i}$:
$$
[H_0,F_{\alpha i}]=-\alpha_{ij}\frac{\partial}{\partial q_j}=i\alpha_{ij}F_j. $$
This is a typical instance of "anomalous conservation" arising from the use of 
a classically indicated Hamiltonian in a quantum context where it is inappropriate---the anomaly 
gets cancelled on adding the anomalous piece $H_\alpha-H_0$ to the "naive Hamiltonian" $H_0$. 
Nontrivial sectors are anomalous in this sense. (For more on the link between PURs and anomalies, see [10]).

The Heisenberg equations for the angle operators $Q_i$ (and $F'_{\alpha i}$),
$$ \dot Q_i = i[H_\alpha, Q_i] = 2F'_{\alpha i},$$
$$ \ddot Q_i = 2 \dot F'_{\alpha i} = 2i[H_\alpha, F'_{\alpha i}] = -4\alpha_{ij}F'_{\alpha j},$$
are easily integrated. Thus the term 'integrable' is justified despite the presence of central 
charges and anomalies.

6.{\it Spectral properties, Weyl group invariance}. The Hamiltonian $H_\alpha$ can be explicitly 
diagonalised as follows. 

As an operator on ${\cal H}_\alpha$, $H_\alpha$ is invariant under $F'_{\alpha i}\rightarrow M_{ij}F'_{\alpha j}=:I_{\alpha i}$ for $M$ a real orthogonal matrix:
$$ H_\alpha = \sum_{i=1}^{2l} (I_{\alpha i})^2, \hskip 5mm [I_{\alpha i}, I_{\alpha j}] = i(MAM^T)_{ij}.$$
The matrix $A$ of central charges is real antisymmetric and hence belongs to the Lie algebra of 
$SO(2l)$. If $T$ is the maximal torus of $SO(2l)$, we can choose $M$ such that $MAM^T=:C$ is in the 
Lie algebra of $T$ (the maximal commuting or Cartan subalgebra)[11]; {\it i.e.,} $C$ has the form 
$$
  C=\left(
    \begin{array}{cc}
           0 & B\\
	   -B & 0
	   \end{array}
	   \right)
$$           
with $B$ a real $l\times l$ diagonal matrix, $B$ = diag($\beta_1,. .,\beta_l$). Redefining 
$I_{\alpha l+i}=:J_{\alpha i},$ we have 
$$
H_\alpha = \sum_{k=1}^l((I_{\alpha k})^2 + (J_{\alpha k})^2), \hskip 5mm [I_{\alpha k},J_{\alpha m}]
=i\delta_{km}\beta_k,
$$
with other commutators zero. Without losing generality, we may therefore assume $H_\alpha$ to have 
the above canonical form which is that for an $l$-dimensional oscillator or $l$ Landau 'electrons' 
with charges proportional to $\{\beta_k\}$. The energy spectrum consists of eigenvalues
$$
E_\nu := E_{\{\nu_k\}} = \sum_{k=1}^l(\nu_k + 1/2)\mid \beta_k \mid ^2
$$
for arbitrary nonnegative integers $\nu_k$. The degeneracy of $E_\nu$ in ${\cal V}_{-\alpha}$ is the number of solutions 
$\{\nu_k\}$ of this equation for fixed $\{\beta_k\}$---for this reason, it is fair to term it 
arithmetic, an especially suitable name when all $\beta_k$ are rational numbers. 

The arithmetic degeneracy has an alternative description 
by means of an invariance property of $H_\alpha$ under the Weyl group of $SO(2l)$[11]. 
 $SO(2l)$ has a subgroup transforming $I_{\alpha k},J_{\alpha k}$ linearly among 
themselves while preserving the canonical structure of their commutators, {\it i.e.,} taking the set 
$\{\beta_k\}$ to some $\{\beta'_k\}$. This group is the normaliser $N$ of $T$ in $SO(2l)$, consisting 
of $M$ such that $MCM^T$ is also in the Cartan subalgebra. Obviously, $H_\alpha$ is invariant 
under $N$. But $N$ itself has a normal subgroup of matrices which take each $C$ to itself, namely 
the centraliser $Z (\cong T)$ of $T$, and $Z$ essentially (upto unitary equivalence) fixes each 
$I_{\alpha k}, J_{\alpha m}$. Therefore, $H_\alpha$ is invariant under the quotient group $N/Z$ 
which by definition is the Weyl group $W$ of $SO(2l)$, alternatively described as the group 
generated by reflections along $l$ basic roots in ${\bf R}^l$[11]. The Weyl group invariance reflects 
the fact that $H_\alpha$ is associated uniquely not to $\alpha$, but to the adjoint orbit of $SO(2l)$ 
through the matrix $A$ in its Lie algebra. The relevance of Weyl groups and root systems to integrability 
thus has a transparent explanation in the quantum context.

It follows that the arithmetic degeneracy of every energy level is the dimension of some representation 
of $W$. It depends critically on the relative values of $\beta_k$; for instance, if $\beta_k/ \beta_m$ 
is irrational for all $k$ and $m$, only the identity representation of $W$ occurs since, in that case, 
for any eigenvalue $E_\nu$ there can be only one solution for $\{\nu_k\}$ in integers. In ${\cal H}_\alpha$ 
itself, there is an additional common infinite (= dim${\cal V}_\alpha$) degeneracy since $H_\alpha$ is 
independent of $F_{\alpha i}$; this reflects the symmetry of the system under the integrability 
group $G$. 

7.{\it Conclusion}. In summary, this paper has delineated a general framework for quantum integrability 
as the most natural quantum generalisation of the classical Liouville-Arnold theory. While incorporating 
normal q-integrable systems with involutive action operators, its scope extends to systems which, though 
involutivity is 'broken' by central charges, remain integrable, in particular exactly diagonalisable. 
The representation-theoretic approach to counting the arithmetic degeneracy appears to be novel and should 
prove useful in other similar physical problems of a Diophantine nature [12]. A fuller treatment, 
including concrete examples, will be taken up elsewhere.

{\it Acknowledgments}. Discussions and correspondence with Radha Balakrishnan, M.Barma, 
N.D.Hari Dass, G.Date, S.Ramanan and R.Sasaki are gratefully 
acknowledged.

\vskip 50mm
\noindent
{\bf References and footnotes}

\noindent
[1] V.I.Arnold, {\it Mathematical Methods of Classical Mechanics}, 2nd ed.
 (Springer, New York, 1989).

\noindent 
[2] F.Calogero, J. Math. Phys. {\bf 10}, 2191 (1969); {\bf 12}, 419 (1971);
B.Sutherland, Phys. Rev. {\bf A4}, 2019 (1971); {\bf A5}, 1372 (1971).

\noindent
[3] M.A.Olshanetsky and A.M.Perelomov, Phys. Reports {\bf 94}, 313 (1983) 
has a systematic study of a variety of such (in our terminology, 
normal) q-integrable models.

\noindent
[4] A great deal of progress is being made currently in this aspect of 
the subject; see E.Corrigan and R.Sasaki, J. Phys.A {\bf 35}, 7017 (2002) 
and work cited there. The role of the Weyl group and root systems 
associated to them is explored in I.Loris and R.Sasaki, J. Phys.A   
{\bf 37}, 211 (2004) and work cited there.

\noindent
[5] Systems which violate this condition have been considered, see P.J.Richens and M.V.Berry, Physica {\bf 2D}, 495 (1981); G.Date, 
M.V.N.Murthy and R.Vathsan, "Classical and Quantum Mechanics of Anyons", 
cond-mat/0302019.

\noindent
[6] A detailed elementary account of the theory of PURs (as it applies 
to quantum mechanics) is given, for general groups, in P.P.Divakaran,
Rev. Math. Phys. {\bf 6}, 167 (1994). For abelian groups, see  
P.P.Divakaran and A.K.Rajagopal, Int. J. Mod. Phys. {\bf 9}, 261 (1995) 
and G.Date and P.P.Divakaran, Ann. Phys. {\bf 309}, 421 (2004).

\noindent
[7] M.S.Raghunathan, Rev. Math. Phys. {\bf 6}, 207 (1994) is essential 
reading for the mathematics of central extensions of groups and 
related topics.

\noindent
[8] V.Bargmann, Ann. Math. {\bf 59}, 1 (1954).

\noindent
[9] For details of this factorisation for $l=1$, see Divakaran and 
Rajagopal, ref.6; this is a special case of a very general property 
that holds for all Heisenberg groups, see D. Mumford with M. Nori 
and P. Norman, {\it Tata Lectures on Theta} III (Birkh\"auser, Boston,
1991).

\noindent
[10] P.P.Divakaran, Phys. Rev. Letters {\bf 79}, 2159 (1997) and ref.6.

\noindent
[11] Any text book on Lie algebras and Lie groups, {\it e.g.,} 
W.Fulton and J.Harris, {\it Representation Theory, A First Course} 
(Springer, New York, 1991).

\noindent
[12] A selected list is: C.Itzykson and J.M.Luck, J. Phys.A {\bf 19}, 211 
(1986); V.Subrahmaniam and M.Barma, J. Phys.A {\bf 24}, 4303 (1991); 
M.N.Tran, M.V.N.Murthy and R.K.Bhaduri, Ann. Phys. {\bf 311}, 204
 (2004).

\end{document}